\documentclass[conference]{IEEEtran}
\IEEEoverridecommandlockouts
\usepackage[T1]{fontenc}
\usepackage{lmodern} 
\usepackage{url}
\usepackage{cite}
\usepackage{booktabs}
\usepackage{amsmath,amssymb,amsfonts}
\usepackage{algorithmic}
\usepackage{graphicx}
\usepackage{textcomp}
\usepackage{xcolor}
\def\BibTeX{{\rm B\kern-.05em{\sc i\kern-.025em b}\kern-.08em
    T\kern-.1667em\lower.7ex\hbox{E}\kern-.125emX}}

\usepackage{tcolorbox}
\usepackage{algorithm}
\usepackage{algorithmic}
\usepackage{pgfplots}
\pgfplotsset{compat=1.18}

\definecolor{viridis1}{HTML}{440154} 
\definecolor{viridis2}{HTML}{3B528B} 
\definecolor{viridis3}{HTML}{21908C} 
\definecolor{viridis4}{HTML}{5DC863} 
\definecolor{viridis5}{HTML}{FDE725} 

\definecolor{viridis6}{HTML}{440154} 
\definecolor{viridis7}{HTML}{443983} 
\definecolor{viridis8}{HTML}{31688E} 
\definecolor{viridis9}{HTML}{21918C} 
\definecolor{viridis10}{HTML}{35B779} 
\definecolor{viridis11}{HTML}{90D743} 
\definecolor{viridis12}{HTML}{FDE725} 

\usepackage{xcolor}

\usepackage{amsmath}     
\usepackage{tikz}
\usetikzlibrary{positioning, arrows.meta}

\begin{document}

\title{Understanding the Perception of Transliterated Offensive Social Media Posts Among Banglish and Hinglish Users}

\author{\IEEEauthorblockN{Dhiman Goswami}
\IEEEauthorblockA{\textit{Computer Science} \\
\textit{George Mason University}\\
Fairfax, VA, USA \\
dgoswam@gmu.edu}
\and
\IEEEauthorblockN{Zhicong Lu}
\IEEEauthorblockA{\textit{Computer Science} \\
\textit{George Mason University}\\
Fairfax, VA, USA \\
zlu6@gmu.edu}
\and
\IEEEauthorblockN{Sanchari Das}
\IEEEauthorblockA{\textit{Information Sciences \& Technology} \\
\textit{George Mason University}\\
Fairfax, VA, USA \\
sdas35@gmu.edu}
}


\maketitle

\begin{abstract}
Effective content moderation depends on understanding not just what harmful content looks like, but how users perceive and respond to it. When offensive language is written in transliterated forms such as Banglish or Hinglish (Bangla or Hindi in Latin script), it may be interpreted very differently than in native scripts, with direct consequences for whether users report, ignore, or recognize it as harmful. Yet little is known about how script choice relates to user perception, or how these perceptions vary across demographic groups. To investigate these perceptions, we conducted a user study with $310$ participants ($193$ Hindi, $60$ Bangla, $57$ Bangla-Hindi bilingual speakers) who evaluated offensive posts in native scripts and transliterations. Participants provided offensiveness scores ($4.03$ for Banglish, $3.86$ for Hinglish) to transliterated posts, yet reporting fell from $20.9\%$ (native - Bangla, Hindi) to $18.2\%$ (transliterated - Banglish, Hinglish). Ignoring remained the dominant response (range $25.38\%-44\%$ overall). Women reported at higher probability than men for native-script ($23.6\%$ vs.\ $17.7\%$) and showed almost identical reporting probability in comparison to men for transliterated posts ($17.5\%$ vs.\ $17.9\%$), and younger participants ($<40$) rated Banglish content as more offensive than older participants ($\geq 40$), (Overall offensive score: $3.99$ vs. $3.78$). These findings identify transliteration as critical blind spot for moderation, underscoring the need for linguistically sensitive approaches to online safety.
\end{abstract}


\section{Introduction}
\label{sec:intro}
Transliteration is the process of writing words from one script into another while aiming to preserve their sound rather than their meaning~\cite{karimi2011machine}. For example, a Bangla or Hindi word can be written in the Latin alphabet instead of its native script. On the other hand, code-mixing or switching refers to the mixture of words of different languages in a single utterance. This practice of transliteration is distinct from code-mixing or switching because it focuses on script conversion and the reproduction of pronunciation~\cite{ho2007code,ali2017study}. Li et al.~\cite{li2009report} defined transliteration as~\textit{\lq\lq...converting names (or text) from one language to another via phonetic translation, preserving pronunciation across distinct scripts or writing systems\rq\rq}. People often use transliteration for practical reasons such as the lack of native-script keyboards, but also as a stylistic or social choice when they want to make text easier to share or more widely accessible. On social media, transliteration has become an increasingly common way of writing. Platforms are optimized for Latin-script keyboards, which encourages users to write in forms such as Banglish (Bangla + English), Hinglish (Hindi + English), Arabizi (Arabic + English), Tenglish (Telugu + English), and Tamlish (Tamil + English), where writing in hybrid scripts reflects both convenience and identity for multilingual communities~\cite{hasan2025code}. For many users these transliterated blends are both a practical solution and an expression of digital culture and belonging. In practice, transliterated posts often combine script conversion with intra-sentential English insertions; we treat a post as transliterated when the native content is rendered in Latin script, regardless of isolated English lexical co-occurrence.

While transliteration facilitates written communication for multilingual users, it also influences how users recognize and respond to harmful content - a dimension that has received far less attention than its computational challenges. Insults, slurs, and coded expressions written in transliterated Latin script often bypass keyword filters and automated moderation~\cite{raihan-etal-2023-offensive,sharma2024thar}, but the more consequential and underexplored question is whether users themselves perceive such content as harmful, and what actions they take when they encounter it. Creative spelling and variation across scripts make transliteration expressive for users, yet these same features create blind spots in how harm is recognized, both by automated systems and by users themselves. As a consequence, vulnerable groups such as adolescents, women, and gender-diverse individuals are disproportionately exposed to harassment that remains hidden in plain sight~\cite{tarihoran2022impact,singh2024mimic}. Current moderation techniques have been trained primarily on monolingual or standard-script data and struggle with the non-standard forms that characterize transliterated communication~\cite{suresh2024comprehensive,yang2025evaluating}. Yet even a perfectly accurate detection system cannot substitute for understanding how real users perceive and respond to harmful transliterated content, since user reporting and engagement behavior are central to how moderation operates in practice. Transliteration therefore represents a double-edged practice: it broadens opportunities for self-expression and online participation while at the same time creating difficulty for moderation and, crucially, shaping whether users recognize and act on the harm they encounter.

Bangla and Hindi speakers provide a relevant and large-scale user base for examining these questions. Together these two languages are spoken by more than $300$ million people, with English widely used as a second language~\cite{ethnologue}. In multilingual contexts, speakers regularly move between these languages in films, music, everyday conversation, and increasingly on social media~\cite{barnali2017code,ahmad2024crossing,noman2019techies,das2019all}. Writing Bangla or Hindi in Latin script has become a common practice, not only in casual digital communication but also within diaspora communities who rely on shared scripts to stay connected~\cite{brandt2018script}. For example:
\begin{quote}
\textit{``Ai jonno ee toh sala ra jei desh a jay sei desher lok ra ogo jutar bari diya bair korey kela pare na abar boro boro kotha bole ei salara''}\\
(Translation: That is why, damn it, wherever these people go, the locals of that country drive them out with shoes, yet they still go around boasting and talking big, these damned ones.) (Context: Sports; Lang: Banglish)
\end{quote}
\begin{quote}
\textit{``Are iski baat ur gadhe ki laat ek jesi h roj palti marti h... iska koi bharosa nhi... main usshe bohot nafrat karti hoon, roj gand nikalti h Shilpa ke liye... ise bas Hina pasand''}\\
(Translation: Her words are like a donkey's kick, changing every day; she cannot be trusted. I hate her so much, every day she causes trouble for Shilpa, and she only supports Hina.) (Context: Reality Show; Lang: Hinglish)
\end{quote}

Despite its prevalence, how users actually perceive and respond to offensive transliterated content remains understudied~\cite{lee2015multilingual}. Prior computational work has incidentally noted that transliterated abusive posts receive lower offensiveness ratings from some users~\cite{raihan-etal-2023-offensive,sengupta2022does} and are deliberately crafted to evade detection~\cite{sharma2024thar}, but these observations are secondary findings rather than the focus of systematic user-centered investigation. Multilingual users' perceptions of transliteration play an important role in shaping research that is both inclusive and practical. By understanding how people perceive harm in transliterated text, and what behavioral responses they exhibit, researchers can uncover challenges, preferences, and cultural factors that automated systems routinely overlook~\cite{humphreys2018automated}. These insights are essential for designing content moderation systems that are not only technically accurate but also grounded in real user experience and sensitive to linguistic diversity. Prioritizing user experience bridges the gap between technology and safer language usage in social media. No prior work has directly and systematically investigated user perception of offensive transliterated content, reporting behavior, and how demographic factors such as age and gender moderate these responses — the gaps our work addresses. To bridge these existing gaps, our study is guided by the following research questions (RQs):

\textbf{RQ1:} How does perceived offensiveness differ in transliterated Bangla or Hindi social media posts?

\textbf{RQ2:} How are transliterated offensive content and user responses such as ignoring or reporting associated?

\textbf{RQ3:} How are age and gender associated with these perceptions and usage towards transliterated offensive content?

Our study with $310$ participants reveals substantial heterogeneity: some participants judged the same post as slightly offensive, while others rated it extremely offensive. This spread indicates that transliteration effects cannot be understood through averages, but depend on individual familiarity, demographic background, and usage intensity. These findings point to transliteration as not merely a script conversion but a factor that interacts with language and user context to shape both the perception and the salience of harmful content. 

Overall our work makes the following key~\textbf{contributions}:

\textbf{(1)} We present a user-centered survey examining associations among language variety, perceived offensiveness, reporting intentions, and demographic or usage characteristics among Bangla, Hindi, and English-speaking participants, with a focus on transliterated offensive content in Banglish and Hinglish alongside native-script baselines in Bangla and Hindi. Across $310$ participants, transliterated posts received offensiveness scores ($4.03$ for Banglish and $3.86$ for Hinglish), which shows transliteration is associated with offensive content perception.

\textbf{(2)} We present a behavioral analysis of user responses to offensive transliterated content. Transliterated posts were associated with weaker reporting behavior than native-script posts ($18.2\%$ vs.\ $20.9\%$), while ignoring remained the dominant response across all varieties ($>$25\% very likely to ignore), with bilingual participants showing the highest passive response rates for transliterated content (33--44\%).

\textbf{(3)} We demonstrate observed differences by age in perceptions for Banglish and by gender in usage (interaction) with transliterated offensive content. Women reported offensive content at higher probability than men for native-script ($23.6\%$ vs.\ $17.7\%$) and showed almost identical reporting probability for transliterated posts ($17.5\%$ vs.\ $17.9\%$), while men showed higher overall posting and interaction rates. Age-based offensiveness perception was observed higher among participants under 40 ($3.99$) than those aged 40 and above ($3.78$), and engagement peaked among middle-aged participants (40--44: $80.77\%$) before declining in older groups.

\section{Background}
Transliteration has evolved from a technical problem of mapping words across scripts into a widespread phenomenon in multilingual online communication with direct consequences for user safety. While early work focused on pronunciation-preserving conversion, recent research emphasizes transliteration as a socio-cultural practice shaped by accessibility, identity, and platform constraints, especially in contexts where Latin keyboards dominate~\cite{spilioti2019transliteration,androutsopoulos2016greeklish}. In everyday use, transliteration lowers entry barriers to digital participation and enables rapid, informal expression across linguistic boundaries. More recently, transliteration has drawn attention for its role in enabling and obscuring offensive content in social media — raising concerns not only about detection system performance, but about whether users themselves recognize and respond to harm when it appears in transliterated form.

\subsection{Transliteration in Offensive Language}
Early computational studies framed transliteration as a core task for multilingual information retrieval and lexicon construction, with the research of Karimi et al.~\cite{karimi2011machine} providing a foundational survey of machine transliteration methods. Sociolinguistic research documented script-switching in bilingual settings~\cite{paradis2000early}, later extending to South Asian social media where Bangla and Hindi are widely written in Latin script~\cite{ali2017study,barnali2017code}.~\cite{hasan2025code} highlighted transliteration as both a practical workaround and a marker of youth culture and accessibility. These strands together positioned transliteration as a linguistic resource, a social identity marker, and an emerging computational challenge.

As social media expanded globally, transliteration surfaced across diverse communities as a vector for offensive content. Arabizi became a prominent case, with studies showing its use to evade moderation~\cite{bensalem2023offensive} and to express dialectal abuse in nonstandard orthography~\cite{boucherit2021offensive}. Similar effects were observed for Spanish~\cite{plaza2021offendes} and Vietnamese~\cite{luu2021large}. The issue gained benchmark status with the shared task by~\cite{zampieri2019predicting}, followed by multilingual OffensEval and SemEval competitions that later incorporated transliterated submissions. Dedicated datasets subsequently revealed systematic performance drops of the models for transliterated offensive language, including Bangla~\cite{raihan-etal-2023-offensive}, Hinglish~\cite{sengupta2022does}, and targeted hate speech~\cite{hartmann2025lost} but users' perception can hardly be defined through experimental analysis. With the rise of LLMs, studies showed cross-script generalization errors leading to mislabeling of offensive transliterated content~\cite{briakou2023searching,fan2025user}, as well as vulnerabilities introduced by prompt engineering in noisy multilingual settings~\cite{alam2025large}, further complicating automated moderation.

Large-scale benchmarks such as SentMix-3L~\cite{raihan-etal-2023-sentmix}, OffMix-3L~\cite{goswami-etal-2023-offmix}, and EmoMix-3L~\cite{raihan-etal-2024-emomix} standardized evaluation across Bangla, Hindi, and English transliterations, while VioLens~\cite{saha2023vio} linked transliterated Bangla content to real-world violence. These resources strengthened empirical evaluation and comparability across systems. However, they are built from the perspective of automated detection and largely overlook the user-side question: how do real people perceive, interpret, and respond to offensive content when it appears in transliterated form?

\subsection{Offensive Language in Social Media}
Research on offensive language in social media has both technical and social challenges. Early models proposed fine-grained classification schemes for detecting harmful content~\cite{razavi2010offensive}, while later work examined how users perceive fairness and justice in moderation decisions, justifying that system legitimacy depends not on accuracy but on user trust~\cite{cai2024content}. Cultural variation in offensiveness perception further complicates moderation, with users in different regions showing markedly different sensitivity thresholds, calling for culturally adaptive systems~\cite{davani2024disentangling}. Platform-level analyses reveal persistent tensions between preventing harmful content and preserving user rights~\cite{schaffner2024community}. Critically, these studies demonstrate that effective moderation cannot be reduced to classification performance alone: user perception, behavioral response, and cultural context all shape whether moderation succeeds in practice. Despite advances in modeling~\cite{raihan2023mixed} and cultural sensitivity~\cite{emran2024masontigers}, most work centers on high-resource Western languages, leaving Bangla and Hindi transliterated contexts underexplored and insufficiently theorized from a user perspective.

\textbf{Transliterated Offensive Language in Social Media:}
Most directly relevant to our work, transliteration shapes how users perceive and engage with harm online. Comparative studies suggest that transliterated abusive posts are not only harder for models to detect but also perceived as less harmful, particularly by younger users~\cite{raihan-etal-2023-offensive,sengupta2022does}, highlighting a socio-technical blind spot that operates on the user side as much as the system side. Similarly,~\cite{sharma2024thar} documented deliberate transliteration strategies used to conceal hate speech from both humans and algorithms, underscoring that the evasion is not purely computational but also relies on how users read and interpret non-standard script. Transliteration also complicates the usability of multilingual chatbots and LLM-powered platforms deployed at scale in terms of understanding non-standard language nuances~\cite{kruk2025banglassist}, with implications for how users seek help or report harm in mixed-script environments. Given that Bangla and Hindi speakers constitute nearly 4\% of the global population~\cite{ethnologue}, transliteration in these languages represents a widespread communicative practice with broad implications for online safety. The literature thus reflects a clear progression: from computational transliteration mapping, to sociolinguistic adoption across multilingual communities, and ultimately to a socio-technical challenge in offensive content moderation. What remains missing is a user-centered investigation that directly measures how people perceive offensive transliterated content, how they respond behaviorally, and how demographic factors such as age and gender moderate these effects — precisely the gaps this work addresses.

\section{Method}
This study employs a survey approach to capture quantitative insights into Bangla and Hindi users' interactions with offensive transliterated content. By combining behavioral measures, perception ratings - the methodology facilitates a broader understanding of participant responses and concerns across Bangla and Hindi transliterated offensive social media posts.  

\subsection{Study Design}

\paragraph{\textbf{Overview of Research Question}}
The survey focused on capturing how participants perceive offensive content when it appears in transliterated forms such as Banglish and Hinglish. Participants were asked about how they encounter and post transliterated content, their comprehension of intended meanings, and the degree of offensiveness they attributed to such posts. They also indicated their likely responses to each post, including ignoring, reacting, commenting, sharing, or reporting, and provided explanations for reporting decisions when applicable. These measures, adapted from established instruments on social media interaction and harmful content perception, enabled systematic comparisons across age and gender groups.  

\paragraph{\textbf{Survey Instrument and Questionnaire}}  
The online survey was designed to capture quantitative measures relevant to participant perceptions of offensive transliterated content. It comprised sections on demographic background, social media usage patterns, frequency of exposure to transliterated posts, and behavioral responses. Participants were also asked questions about comprehension, privacy risks and offensive content detection. They were asked to provide transliterated offensive social media post and rated their offensiveness. Standardized Likert-scale~\cite{joshi2015Likert} items were employed throughout to ensure consistency in assessing exposure, perceived offensiveness, acceptability, safety, and reporting behavior.

The survey included items on exposure to and comprehension of transliterated content, perceived harmfulness, behavioral responses, privacy and security concerns. The question types include multiple choices, multiple answers, matrix table, and open ended. General posting and interaction questions were asked once. After that each participant evaluated one post per language based on their declared proficiency: Bangla speakers saw one Bangla, one Banglish, and one English post; Hindi speakers saw one Hindi, one Hinglish, and one English post; and Bangla-Hindi speakers saw posts of all five languages. Posts were drawn from five publicly available datasets, one per language variety, and each carried an existing label which were not shared with the participants. English posts were included as a baseline condition, given that both transliterated varieties use Latin script. Post assignment was randomized such that each post was evaluated by an approximately equal number of participants (some responses were discarded after data collection, yielding the final sample of 310). Participants saw only the posts corresponding to their selected languages and received no translations. Those who declared proficiency in none of the target languages were redirected to the end of the survey. The survey questionnaire is available in the Appendix~\ref{app:ques}. The study design overview is shown in Table~\ref{tab:study_design}. For each participant, one post per language condition was presented based on their language proficiency, and each unique post was shown to a maximum of five participants to obtain independent annotations while limiting repeated exposure effects. The post pool was constructed to include balanced native-script and transliterated variants, with posts examined for potential confounding factors such as target entity, topic, length, profanity usage, and type of harm. All posts were reviewed by two native speakers to ensure linguistic validity and transliteration quality and the disagreements were solved by discussion. During the survey, language conditions and post order were randomized, and participants were not informed of the specific study hypotheses to reduce potential bias in evaluating offensiveness, reporting, and engagement intentions.

\paragraph{\textbf{Dataset Description}}  
We used offensive posts from benchmark datasets in five languages to enable the comparison between native and transliterated ones. These datasets are originally curated from multiple social media posts (Meta, Reddit, X) and adhere to the social media data collection standard. The dataset included:  
\begin{itemize}
    \item \textbf{English:} 120 posts from OLID~\cite{zampieri2019predicting},  
    \item \textbf{Bangla:} 80 posts from the Violence Inciting Text Detection dataset~\cite{saha2023vio},  
    \item \textbf{Hindi:} 80 posts from the Hostility Detection dataset~\cite{bhardwaj2020hostility},  
    \item \textbf{Banglish:} 80 posts from TB-OLID~\cite{raihan-etal-2023-offensive},  
    \item \textbf{Hinglish:} 80 posts from Hinglish Hate Detection dataset~\cite{sengupta2022does}.  
\end{itemize}

There is no single dataset that contains data in a specific language along with translations into two other languages and transliterated variants of each. Since transliteration lacks a standardized form, instead of augmenting a single dataset to create five variants, we chose five different offensive-language datasets selected for approximately similar offensive severity. Posts were labeled transliterated if they (i) originated from a dataset explicitly curated as transliterated Bangla or Hindi, and (ii) native-speaker reviewers confirmed that the majority of content-bearing tokens were Romanized Bangla/Hindi rather than English, distinguishing transliteration from code-mixing. Posts with incidental English words were retained provided the dominant lexical and grammatical structure remained Bangla/Hindi; posts where English carried the primary offensive content were excluded. This classification was applied during the native-speaker linguistic validity review.

\paragraph{\textbf{Participant Recruitment}} We recruited participants from Prolific~\cite{prolific}.
Given the focus on transliterated Bangla-Hindi-English offensive content, participants were required to know at least two of these languages, including English. Screening questions ensured compliance with this requirement. To reduce gender bias and examine gender-related perceptions, we implemented quotas for a balanced gender ratio. All participants were at least 18 years old.  A total of $416$ individuals initially took part. Exclusions were applied for lack of consent ($9$), absence of relevant language knowledge ($14$), duplicate attempts ($23$), and incomplete responses ($60$). The final sample included $310$ participants. Among them 140 were identified as men, $166$ were identified as women, and $4$ preferred not to disclose gender. With respect to language proficiency, $193$ were Hindi-English speakers, $60$ were Bangla-English speakers, and 57 were proficient in all three languages. We used attention check questions where participants were asked to write a native and a transliterated sentence according to their language choice. This ensures the balance between self-reported language proficiency and script literacy.

\paragraph{\textbf{Sample Size Considerations}}

The study included 310 participants distributed across three language proficiency groups: Hindi-English ($n=193$), Bangla-English ($n=60$), and Bangla-Hindi-English ($n=57$). Each participant evaluated one post from each language available to their proficiency group, while each post was evaluated by multiple participants. Because analytical power calculations based on independent-group comparisons do not account for this repeated-measures structure, we do not report post hoc power estimates. Such power analyses have been shown to provide limited statistical insight beyond the observed effect estimates~\cite{hoenig2001abuse}. Throughout the manuscript, we report observed proportions with 95\% confidence intervals as descriptive summaries~\cite{cumming2014new}. The implications of using aggregate proportion-based analyses for a repeated-measures design are discussed in the limitations.

\paragraph{\textbf{Survey Platform Details}}
The survey was deployed on the George Mason University's Qualtrics platform in compliance with the ethical protocol. Multiple submissions were prevented using duplicate detection and bot-detection mechanisms. Search engine indexing was blocked, and IP addresses, locations, and contact details were not collected. To further ensure data quality, we used reCAPTCHA in Qualtrics (average score = $0.986$, range = $0.2–1.0$), included two attention checks midway through the survey, and manually verified participant responses to these checks. These minimized the risk of automated responses and inattentive participation.  

\paragraph{\textbf{Approval of Ethical Review Board}}  
The study was approved by the George Mason University Institutional Review Board - IRB \textbf{(STUDY00000572)} after a comprehensive review. Participants provided informed consent, were informed of potential psychological discomfort from offensive content, and were reminded of their right to withdraw. Minimal demographic data were collected (age, gender, language), no PII was recorded, and recruitment was conducted via Prolific~\cite{prolific}. We used Prolific as we can control the specific group of participants who are eligible for this survey in terms of language proficiency which will ensure quality data collection. This kind of approach was previously employed for prior user studies~\cite{cutler2025crowdsourced}. The survey was hosted on the George Mason University's Qualtrics platform~\cite{qualtrics} with secure data storage, and the full survey and sample posts were reviewed before deployment approval. We used Qualtrics as the questionnaire platform as it is also a well recognized platform for this kind of study~\cite{velykoivanenko2024designing}.

\paragraph{\textbf{Payment Details}}  
Participants received payment upon successful completion of the survey verified by a completion code. The base payment amount is equivalent to 1.50 U.S. dollars and the additional bonus amount for eligible participants (Bangla, Hindi bilingual speakers) is equivalent to 1.0 U.S. dollar.  Average completion time for eligible participants was 15 minutes and 45 seconds.

\begin{table*}[ht]
\centering
\caption{Study design showing participant groups, language-specific post assignments, and outcome measures.}
\label{tab:study_design}
\resizebox{\textwidth}{!}{%
\begin{tabular}{lllcl}
\toprule
\textbf{Participant Group} &
\textbf{Languages Evaluated} &
\textbf{Posts per Participant} &
\textbf{Participants per Post} &
\textbf{Measures Collected} \\
\midrule
Bangla-English &
Bangla, Banglish, English &
1 post per language &
$\leq$5 participants &
Offensiveness, Reporting, Engagement \\

Hindi-English &
Hindi, Hinglish, English &
1 post per language &
$\leq$5 participants &
Offensiveness, Reporting, Engagement \\

Bangla-Hindi-English &
Bangla, Hindi, Banglish, Hinglish, English &
1 post per language &
$\leq$5 participants &
Offensiveness, Reporting, Engagement \\
\bottomrule
\end{tabular}
}
\end{table*}

\subsection{Statistical Analysis}





To address our RQs, first, we examine age and gender-based offensiveness perception, then action and interaction behaviors by comparing production and consumption patterns across language groups, followed by demographic analyses to assess gender and age-related differences in engagement. We quantify the likelihood of actions such as ignoring, commenting, sharing, or reporting, and compare these across languages. Incomplete or partially submitted responses were discarded before analysis. Proportions are computed as the percentage of participants selecting each response, and pairwise comparisons are reported descriptively as proportions with 95\% confidence intervals.




For Likert-scale items, responses in the \textit{Likely} and \textit{Very Likely} categories were pooled for reporting purposes. For Offensiveness, posts marked as \textit{Clearly} to \textit{Extremely Offensive} were pooled for analysis.


\section{Results}
To understand participants' perception of transliterated offensive content in social media, we present the results of our analysis of the surveyed data organized around our three research questions.

\subsection{RQ1: Transliteration and Perceived Offensiveness}

\begin{table}[ht]
\centering
\caption{Pairwise comparison of reporting across languages and varieties for Bangla, Hindi, and Bangla-Hindi bilingual speakers. \textbf{Bold $p$-values indicate significance after Bonferroni correction within each language group.} $p_1$ and $p_2$ denote the fractions of participants in the corresponding comparison groups.}
\label{tab:report_combined}
\resizebox{\linewidth}{!}{%
\begin{tabular}{l|ccccc}
\hline
Comparison & $p_1$ & $p_2$ & 95\% CI & $z$ & $p$-value \\
\hline
\multicolumn{6}{c}{\textbf{Bangla speakers} ($\alpha=0.0167$)} \\
\hline
Bangla vs English   & 0.3667 & 0.1500 & (0.05, 0.39) & 2.68 & \textbf{0.007} \\
Bangla vs Banglish  & 0.3667 & 0.1167 & (0.09, 0.41) & 3.14 & \textbf{0.0017} \\
English vs Banglish & 0.1500 & 0.1167 & (-0.14, 0.21) & 0.53 & 0.60 \\
\hline
\multicolumn{6}{c}{\textbf{Hindi speakers} ($\alpha=0.0167$)} \\
\hline
English vs Hindi    & 0.197 & 0.171 & (-0.09, 0.14) & 0.43 & 0.67 \\
English vs Hinglish & 0.197 & 0.150 & (-0.09, 0.18) & 0.69 & 0.49 \\
Hindi vs Hinglish   & 0.171 & 0.150 & (-0.10, 0.14) & 0.34 & 0.74 \\
\hline
\multicolumn{6}{c}{\textbf{Bangla--Hindi bilingual speakers} ($\alpha=0.005$)} \\
\hline
Bangla vs English   & 0.1228 & 0.2105 & (-0.23, 0.06) & -1.06 & 0.29 \\
Bangla vs Hindi     & 0.1228 & 0.2632 & (-0.31, 0.03) & -1.76 & 0.078 \\
Bangla vs Banglish  & 0.1228 & 0.3333 & (-0.38, -0.04) & -2.49 & 0.013 \\
Bangla vs Hinglish  & 0.1228 & 0.2105 & (-0.23, 0.06) & -1.06 & 0.29 \\
English vs Hindi    & 0.2105 & 0.2632 & (-0.25, 0.14) & -0.64 & 0.52 \\
English vs Banglish & 0.2105 & 0.3333 & (-0.30, 0.05) & -1.47 & 0.14 \\
English vs Hinglish & 0.2105 & 0.2105 & (-0.18, 0.18) & 0.00 & 1.00 \\
Hindi vs Banglish   & 0.2632 & 0.3333 & (-0.24, 0.10) & -0.84 & 0.40 \\
Hindi vs Hinglish   & 0.2632 & 0.2105 & (-0.14, 0.25) & 0.64 & 0.52 \\
Banglish vs Hinglish& 0.3333 & 0.2105 & (-0.05, 0.30) & 1.47 & 0.14 \\
\hline
\end{tabular}%
}
\end{table}

To understand how transliteration shapes the perception of offensiveness, we asked the participants to provide rating to the Banglish or Hinglish (or both, for bilingual speakers) transliterated offensive social media post on a five-point Likert scale (1–5) ranging from slight, moderate, clear, very, to extreme offensiveness.

For Banglish posts, a total of 48 men and 69 women provided scores, with mean offensiveness scores of 4.09 and 3.98, respectively. For Hinglish posts, 117 men and 129 women participated, with mean scores of 3.8 and 3.91, respectively. Overall, mean offensive score of Banglish is $(48 \times 4.09+69 \times 3.98)/(48+69) = 4.03$ and Hinglish is $(117 \times 3.8+129 \times 3.91)/(117+129) = 3.86$. These results indicate that both Banglish and Hinglish transliterated content are perceived as clear to very offensive as per Likert scale derived from separate source datasets per language.

\subsection{RQ2: User Response towards Transliterated Offensive Content}

\begin{table*}[ht]
\centering
\caption{Comparison of perceived offensiveness by gender and age for Banglish and Hinglish. Bold $p$-values indicate statistical significance after Bonferroni correction ($\alpha=0.025$ within each language). $p_1$ and $p_2$ denote the proportions of participants in the compared groups.}
\label{tab:offensiveness_age_gender}
\resizebox{\linewidth}{!}{%
\begin{tabular}{l|ccccc|ccccc}
\hline
Comparison & \multicolumn{5}{c|}{Banglish} & \multicolumn{5}{c}{Hinglish} \\
\cline{2-11}
 & $p_1$ & $p_2$ & 95\% CI & $z$ & $p$-value 
 & $p_1$ & $p_2$ & 95\% CI & $z$ & $p$-value \\
\hline
\multicolumn{11}{c}{\textbf{By Gender}} \\
\hline
Men vs Women & 0.729 & 0.623 & (-0.023, 0.235) & 1.60 & 0.110 
             & 0.675 & 0.643 & (-0.099, 0.163) & 0.48 & 0.63 \\
\hline
\multicolumn{11}{c}{\textbf{By Age}} \\
\hline
Below 40 vs 40 and above & 0.709 & 0.500 & (0.077, 0.341) & 3.10 & \textbf{0.002} 
                         & 0.659 & 0.717 & (-0.186, 0.070) & -0.89 & 0.373 \\
\hline
\end{tabular}
}
\end{table*}

Understanding participants' behavioral responses to offensive content across languages and scripts is central to designing effective moderation systems~\cite{zhou2024understanding,niu2023survey}. We examine two primary dimensions: ignoring behavior and reporting behavior, analyzed across native-script and transliterated varieties for Bangla, Hindi, and bilingual participants.

\subsubsection{\textbf{Ignoring Behavior}}

Across all language groups and script varieties, ignoring was the most frequently chosen response to offensive posts. Among Bangla speakers ($60$), 41.67\% of participants were likely or very likely to ignore Bangla posts, with similarly high ignoring rates for English (41.67\%) and Banglish posts (43.35\%), the highest of the three. Very likely engagement in active actions such as commenting, sharing, or reporting remained consistently low (5–10\%) across all post types, indicating a predominantly passive response pattern. These results show that even in the native language, Bangla speakers rarely reach the ``very likely'' threshold for active interactions.

Hindi speakers ($193$) followed a similar pattern. For English posts, 31.08\% were likely or very likely to ignore, with very low percentages for reacting (9.80\%), commenting (7.77\%), sharing (6.73\%), or reporting (9.84\%). Hindi posts showed slightly lower ignoring rates (25.38\%) in the likely or very likely category, suggesting marginally higher engagement potential, though active interactions remained minimal. Hinglish posts again exhibited high ignoring rates (31.03\%) with low very likely engagement in other actions.

Bilingual participants (proficient in both Bangla and Hindi) ($57$) displayed the most balanced interaction patterns. While a notable fraction were likely or very likely to ignore offensive posts (26–36\% for Bangla, English, and Hindi posts), substantial engagement was also observed in the Neutral and Likely categories across reacting, commenting, sharing, and reporting. Banglish and Hinglish posts showed slightly elevated ignoring behavior (33–44\%), yet participants continued to engage significantly in Neutral categories, reflecting their adaptability to mixed-language content.

\subsubsection{\textbf{Reporting Behavior}}

Despite the overall trend toward passive engagement, reporting patterns reveal meaningful differences across script varieties. Overall, $24.7\%$ (29/(60+57)) of Bangla posts were reported (by Bangla and Bangla-Hindi participants) compared to $22.2\%$ (26/(60+57)) of Banglish posts (by Bangla and Bangla-Hindi participants), and $19.2\%$ (48/(193+57)) of Hindi posts (by Hindi and Bangla-Hindi participants) compared to $16.4\%$ (41/(193+57)) of Hinglish posts (by Hindi and Bangla-Hindi participants), indicating a consistent but modest reduction in reporting when content is transliterated. Overall 20.9\% ((29+48)/(60+57+193+57)) of native (Bangla and Hindi) and 18.2\% ((26+41)/(60+57+193+57)) of transliterated posts are reported by the participants. Bangla speakers reported Bangla posts at higher rates than both English and Banglish posts, whereas reporting rates were similar among Hindi and bilingual speakers across script varieties, as detailed in Table~\ref{tab:report_combined}.


\subsection{RQ3: Age and Gender Influence on Perceptions and Transliteration Usage}

\begin{table*}[ht]
\centering
\caption{Comparison of self-reported general transliteration usage by gender and age of the participants, not their responses to the offensive posts presented in the survey. Bold $p$-values indicate statistical significance after Bonferroni correction ($\alpha=0.025$ within each outcome). $p_1$ and $p_2$ denote the proportions of participants in the compared groups.}
\label{tab:gender_age_post_interact}
\resizebox{\linewidth}{!}{%
\begin{tabular}{l|ccccc|ccccc}
\hline
Comparison & \multicolumn{5}{c|}{Posting} & \multicolumn{5}{c}{Interaction} \\
\cline{2-11}
 & $p_1$ & $p_2$ & 95\% CI & $z$ & $p$-value 
 & $p_1$ & $p_2$ & 95\% CI & $z$ & $p$-value \\
\hline
Men vs Women & 0.700 & 0.602 & (-0.072, 0.268) & 1.13 & 0.260 
             & 0.829 & 0.607 & (0.155, 0.501) & 3.34 & \textbf{0.0008} \\
\hline
Below 40 vs 40 and above & 0.621 & 0.664 & (-0.176, 0.090) & -0.634 & 0.526 
                         & 0.368 & 0.436 & (-0.204, 0.068) & -0.981 & 0.326 \\
\hline
\end{tabular}
}
\end{table*}

To examine whether age and gender impact perceptions of and usage to transliterated offensive content, we conducted a series of statistical comparisons across these demographic dimensions.

\subsubsection{\textbf{Gender-based Offensiveness Perception}}
Women (weighted average score: 3.95) and men (3.89) both perceived Banglish and Hinglish content as strongly offensive. This near-identical scoring suggests that gender has minimal influence on the perception of online offensive content, and sensitivity to harmful transliterated language is largely consistent across both groups. Men and women showed similar proportions rating Banglish or Hinglish posts as clearly, very, or extremely offensive, as shown in Table~\ref{tab:offensiveness_age_gender}.


\subsubsection{\textbf{Gender-based Transliteration Usage}}
Women ($166$) exhibited a higher probability of reporting native-script posts than men ($140$) ($23.6\%$ vs.\ $17.7\%$). In contrast, the probability of reporting transliterated posts was nearly identical across genders ($17.5\%$ for women vs.\ $17.9\%$ for men), suggesting that transliteration substantially attenuates the gender difference in reporting behavior.

Regarding general transliteration usage, a higher proportion of men reported both posting and interacting with transliterated content (70\%) compared to women (59.6\%), reflecting greater overall use of transliteration among male participants. Women were more likely than men to report interacting with transliterated content without posting it (28.3\% vs.\ 17.1\%), suggesting a tendency toward passive engagement with transliterated content. Posting transliterated content without interacting with it was rare for both genders, and a small proportion of participants reported neither posting nor interacting with transliterated content, at similar rates for men (12.9\%) and women (11.4\%). Self-reported transliteration posting was similar between men and women, while self-reported interaction showed a larger gap between the two groups, as shown in Table~\ref{tab:gender_age_post_interact}.


\subsubsection{\textbf{Age-based Offensiveness Perception}}
The perceived offensiveness of transliterated content varied across age groups. For Banglish, the 18–24 group showed mixed responses with no single category dominating, while those aged 25–29 and 35–39 were more likely to assign higher severity, with a strong concentration of extreme scores. The 30–34 group leaned toward mid-level offensiveness, and middle-aged groups (40–54) demonstrated diverse perceptions. Older participants (55–64) provided fewer responses, with the 60–64 group skewing heavily toward a single category, likely due to low sample size. For Hinglish, younger participants (18–24) leaned toward mid-to-high severity, with clear and very offensive judgments. The late 20's group clustered around clear and very offensive scores, while the mid-30s cohort showed a strong concentration of very and extreme responses. Middle-aged participants (40–54) displayed broad variation without a dominant category, and older participants (55–64) showed similar sparsity, with the 60–64 group overwhelmingly labeling content as very offensive.

For statistical analysis we divided the participants into two groups: younger adults (age $< 40$, 244 participants) and older adults (age $\geq 40$, 66 participants). While no standardized age threshold exists in prior work for distinguishing early versus later social media adopters, we set the age cutoff at 40 to separate participants who were exposed to social media before or during their teenage years from those who encountered it only in adulthood. Since mainstream social media adoption began in the mid-to-late 2000s, participants under 40 would have been teenagers or younger at that time, while those 40 and older would have already passed their teenage years distinguishing early, formative exposure from later, adopted use. Participants under 40 gave a weighted average offensiveness score of $3.99$, reflecting moderate-to-strong sensitivity. Participants aged 40 and above scored slightly lower at $3.78$, yet still rated content as clearly offensive. Since age-stratified subgroups spanning five-year intervals contain too few participants to satisfy the normal analysis conditions, participants were consolidated into two broader age groups: below 40 and 40 or above. A gap between these two age groups was observed in the acceptance of Banglish offensiveness, while the two groups showed similar proportions for Hinglish, as shown in Table~\ref{tab:offensiveness_age_gender}.


\subsubsection{\textbf{Age-based Transliteration Usage}}
General transliteration usage also varied across age groups. Participants aged 25--29 and 40--44 reported the highest proportions of both posting and interacting with transliterated content (68.35\% and 80.77\%, respectively), suggesting greater transliteration usage among young and early middle-aged adults. The 55--59 age group also reported higher levels of posting and interaction, though the sample size is likely very small. Older participants (60--64) reported the lowest levels of transliteration usage, with a notable fraction reporting neither posting nor interaction. Across almost all age groups, participants were more likely to report interacting with transliterated content than posting it. The below 40 and 40 and above groups showed similar proportions for both self-reported posting and interaction, as shown in Table~\ref{tab:gender_age_post_interact}.


\section{Discussion}
Transliteration functions as a socio-technical variable rather than a superficial script change. It shapes how users evaluate and act on offensive content and undermines automated moderation by introducing spelling variability, mixed-script tokens, and contextual ambiguity~\cite{alharbi2024transliteration,magdy2025should}. These properties create systematic blind spots for detection and triage, linking human perception with algorithmic error of moderation.

On average, transliterated posts had offensiveness scores of $4.03$ for Banglish and $3.86$ for Hinglish in Likert scale. Yet this overall difference conceals substantial heterogeneity in offensiveness perception. 

Older participants ($\geq 40$) tended to down-weight transliterated harms. This group showed greater tolerance for creative spelling and informal transliteration practices, reflecting a greater willingness to interpret such language within its broader social and communicative context. By contrast, younger participants ($<40$) judged the same posts as more offensive. For them, transliteration often heightened rather than softened the perceived severity of offensive language. This divergence states age as an indicator in the interpretation of transliterated harms.

Gender patterns exhibited a degree of symmetry. Women participants exhibited a reporting probability similar to that of men for transliterated content, with a $17.5\%$ probability of reporting among women compared to $17.9\%$ among men. Even though the reporting probability is not significant, this percentage still emphasizes the cumulative impact of repeated exposure to gendered abuse written in transliterated form. While men showed higher overall engagement with offensive content, their actions skewed toward passive strategies (e.g., viewing or reacting without reporting).

Gender and age also influenced participants' perceptions and engagement with transliterated offensive content. Younger participants (age $<40$) offensiveness ratings differ than older participants (3.99 vs.\ 3.78), while women rated content as marginally more offensive than men (3.95 vs.\ 3.89) in Likert scale. Behavioral differences revealed that men were more likely to both post and interact with (70.0\% vs.\ 59.6\%), whereas women more frequently engaged only through interaction (28.3\% vs.\ 17.1\%).

\begin{tcolorbox}[
    colback=blue!5!white,
    colframe=blue!60!black,
    colbacktitle=blue!60!black,
    coltitle=white,
    title={\textbf{Key Findings}},
    fonttitle=\bfseries,
    arc=1mm,
    boxrule=1pt,
    left=1pt,
    right=1pt,
    top=1pt,
    bottom=1pt,
]
\textbf{RQ1: Perceived Offensiveness}
\begin{itemize}
    \item \textbf{Men/Women} 4.09/3.98 (Banglish), 3.80/3.91 (Hinglish)
    
\end{itemize}
\textbf{RQ2: Users' Response}
\begin{itemize}
    \item \textbf{Bangla Speakers:} (very likely): 41.67\% (Bangla), 41.67\% (English), 43.35\% (Banglish) - Ignoring
    \item \textbf{Hindi Speakers:} (very likely): 25.38\% (Hindi), 31.03\%(Hinglish); active actions $<$10\% across all - Ignoring
    \item \textbf{Bilingual Speakers:} 26--36\% (native) vs.\ 33--44\% (transliterated) - Ignoring
    \item \textbf{Reporting:} 24.7\% (Bangla) vs.\ 22.2\% (Banglish); 19.2\% (Hindi) vs.\ 16.4\% (Hinglish)
\end{itemize}
\textbf{RQ3: Gender/Age-based Behavioral Association}
\begin{itemize}
    \item \textbf{Offensiveness:} 3.99 (Age $<$40) vs.\ 3.78 (Age $\geq$40); 3.95 (Women) vs.\ 3.89 (Men)
    \item \textbf{Gender:} Men post+interact 70\% vs.\ Women 59.6\%; Women interact-only 28.3\% vs.\ Men 17.1\%
    \item \textbf{Age:} peaks at 25--29 (68.35\%) and 40--44 (80.77\%), lowest at 60--64
\end{itemize}
\end{tcolorbox}

Thus, transliteration does not produce a uniform perceptual shift. Rather, its impact is contingent on demographic characteristics, everyday communication practices, and the norms of the platforms where it appears. What seems like a simple change of script on the surface operates as a layered phenomenon, encoding deeper differences in how users interpret offensive language and in the actions they take in response~\cite{boucher2022bad}. Transliteration therefore functions as both a linguistic transformation and a social signal that modifies the dynamics of recognition, tolerance, and intervention~\cite{dias2016language}.

\section{Implications}
Based on our findings about participants' perceptions of transliterated offensive posts, we outline several hypotheses and directions for future design research aimed at improving platform safety and content moderation in multilingual contexts.

\subsubsection{Variant-Aware User Controls}
Our findings suggest that transliteration influences reporting tendencies and that many users default to ignoring offensive content. Future research could investigate whether variant-aware user controls improve users' ability to manage exposure to transliterated offensive content. For example, studies could evaluate \textit{filtering rules} that allow users to block transliterated posts based on perceived severity, together with \textit{sensitivity sliders} that adjust tolerance toward specific linguistic cues~\cite{dave2021social}. Future work could also examine whether previewing the effects of these settings helps users better understand and trust content filtering decisions.

\subsubsection{Community-Engaged Review Workflows and Human-in-the-Loop Moderation}
Because reporting rates were lower for transliterated than native-script posts, future work could investigate whether moderation systems that combine automated transliteration-aware detection with community-engaged review workflows improve the identification of transliterated abuse~\cite{wu2022survey}. One possible direction is to evaluate tiered review pipelines in which posts are routed according to estimated offensiveness and reviewed by multilingual community members or expert moderators depending on confidence and severity.

\subsubsection{Context-Aware User Education}
Our results indicate demographic differences in perceptions of transliterated offensiveness. Future research could examine whether contextual educational interventions, such as inline explanations of transliteration patterns, localized safety guidance, or examples of evasive language, improve users' ability to recognize and respond to offensive transliterated content~\cite{al2010transliteration}. Longitudinal studies could further evaluate whether user feedback helps these educational resources adapt to evolving transliteration practices.

\subsubsection{Transparency}
Future work could investigate whether providing explanations for moderation decisions involving transliterated content improves user understanding, trust, and reporting behavior. Possible approaches include presenting cue-level explanations, such as script or language-variant information, together with moderation outcomes and user controls~\cite{juneja2020through}. Controlled user studies could assess whether such explanations influence users' perceptions of moderation fairness and effectiveness.

\subsubsection{Participatory Methods, Evaluation, and Ethics-in-Use}
Our findings suggest that perceptions of transliterated offensiveness vary across demographic groups. Future design research could explore participatory approaches, including co-design workshops and cognitive walkthroughs with diverse user populations, to better understand how different communities perceive and respond to transliterated offensive content~\cite{davis2021low}. Such studies may help inform moderation policies and interface designs that better reflect the needs and expectations of multilingual users.

\section{Limitations and Future Work}
We provide insight into how transliteration shapes perceptions of offensiveness; however, our study carries several limitations. Our sample was restricted to participants recruited through Prolific, which was chosen for the quality of survey data it provides, similar to prior studies~\cite{douglas2023data}. At the time of data collection, Prolific could hardly capture the perspectives of gender-diverse participants, who were also underrepresented due to small subgroup sizes, as were older age groups. Although participants contributed repeated ratings and each post was evaluated by multiple participants, our analyses summarized aggregate group differences using descriptive proportions and confidence intervals to provide direct and interpretable comparisons of observed reporting rates. Consequently, participant and post-level variability was not modeled explicitly, and future work should validate these findings using generalized linear mixed-effects models with random effects for participants and posts. Post selection across five publicly available datasets may introduce severity imbalances, and findings may not generalize beyond Bangla, Hindi, and English speakers, nor fully capture cultural and regional variation within these language communities. Additionally, exposing participants to offensive content carries ethical considerations; we mitigated this through informed consent and the option to withdraw at any point. Moving forward, we aim to incorporate data from more inclusive gender identities, conduct cross-platform comparisons, and design behavioral experiments in naturalistic environments. We also plan to extend our analysis to additional languages to enable cross-lingual comparisons and investigate how educational backgrounds, particularly the medium of instruction, influence transliteration practices and perceptions of offensive content. Such analyses can provide insights into the emergence of transliterated varieties such as Banglish and Hinglish and their role in shaping online communication.

\section{Conclusion}
Social media is central to everyday communication, and in multilingual communities this communication often occurs through transliteration. Because transliterated content can cross linguistic boundaries while preserving the meaning of offensive language, understanding how users perceive and respond to it is important for designing effective moderation systems. We presented a user-centered study of $310$ multilingual participants examining perceptions and reported responses to offensive content written in Bangla, Hindi, Banglish, and Hinglish. Our findings provide three key insights. First, participants perceived transliterated offensive content as very offensive across both Banglish and Hinglish posts. Second, perceptions of transliterated offensiveness varied across demographic groups, with observed age-based differences in offensiveness ratings for Banglish. Third, participants' reported responses were predominantly passive, with ignoring being the most common action and reporting being relatively infrequent. Together, these findings highlight that transliterated offensive content is perceived differently across users and language varieties, underscoring the importance of incorporating user-centered and multilingual considerations into the design of content moderation systems.

\bibliographystyle{plain}
\bibliography{custom.bib}

\section*{Data Samples}

\begin{quote}
\raisebox{-2.5pt}{\includegraphics{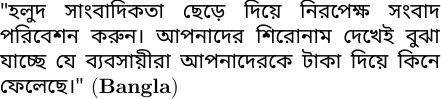}}\\
(Translation: Stop yellow journalism and deliver unbiased news. Just by looking at your headlines, it is clear that businessmen have bought you off with money.) 
\end{quote}

\begin{quote}
\raisebox{-2.5pt}{\includegraphics{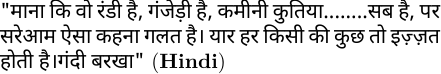}}\\
(Translation: It's true that she is a prostitute, bald, a wicked woman... all of that, but saying it openly like this is wrong. Everyone has some dignity, after all. Dirty Barkha) 
\end{quote}

\begin{quote}
\textit{"Rip lmao who knows that's why I'm trying so hard not to expose my shits since they can't never relate anyway"} (\textbf{English})
\end{quote}

\begin{quote}
\textit{``Ai jonno ee toh sala ra jei desh a jay sei desher lok ra ogo jutar bari diya bair korey kela pare na abar boro boro kotha bole ei salara''} ( \textbf{Banglish})\\
(Translation: That is why, damn it, wherever these people go, the locals of that country drive them out with shoes, yet they still go around boasting and talking big, these damned ones.) 
\end{quote}

\begin{quote}
\textit{``Are iski baat ur gadhe ki laat ek jesi h roj palti marti h... iska koi bharosa nhi... main usshe bohot nafrat karti hoon, roj gand nikalti h Shilpa ke liye... ise bas Hina pasand''}  (\textbf{Hinglish})\\
(Translation: Her words are like a donkey's kick, changing every day; she cannot be trusted. I hate her so much, every day she causes trouble for Shilpa, and she only supports Hina.)
\end{quote}

\section{Survey Questionnaire}
\label{app:ques}

\subsection{Consent Form}
\begin{itemize}
\item Do you agree to the consent form?
\begin{itemize}
    \item[$\bigcirc$] Yes \hspace{1cm} $\bigcirc$ No
\end{itemize}
\end{itemize}

\subsection{Demographics and Background}
\begin{itemize}
\item What is your age group?
\begin{itemize}
    \item[$\bigcirc$] less than 18
    \item[$\bigcirc$] 18--24
    \item[$\bigcirc$] 25--29
    \item[$\bigcirc$] 30--34
    \item[$\bigcirc$] 35--39
    \item[$\bigcirc$] 40--44
    \item[$\bigcirc$] 45--49
    \item[$\bigcirc$] 50--54
    \item[$\bigcirc$] 55--59
    \item[$\bigcirc$] 60--64
    \item[$\bigcirc$] 65 or above
\end{itemize}

\item Which gender do you mostly identify with?
\begin{itemize}
    \item[$\bigcirc$] Men
    \item[$\bigcirc$] Women
    \item[$\bigcirc$] Non-Binary
    \item[$\bigcirc$] Prefer not to say
    \item[$\bigcirc$] Prefer to Self Describe (Please specify)
\end{itemize}

\item Which language(s) do you understand apart from English? (Select all that apply)
\begin{itemize}
    \item Bangla
    \item Hindi
    \item None of the above
\end{itemize}
\end{itemize}

\subsection{Social Media Interaction}
\begin{itemize}
\item How much time do you spend daily on average on social media?

\begin{table}[htbp]
\centering
\resizebox{\linewidth}{!}{%
\begin{tabular}{l | c c c c c c c}
\hline
\textbf{Platform} &  < 30 min & 30 min--1 hr & 1--2 hr & 2--3 hr & 3--4 hr & > 4 hr & Don't use \\
\hline
Facebook & $\bigcirc$ & $\bigcirc$ & $\bigcirc$ & $\bigcirc$ & $\bigcirc$ & $\bigcirc$ & $\bigcirc$ \\
YouTube & $\bigcirc$ & $\bigcirc$ & $\bigcirc$ & $\bigcirc$ & $\bigcirc$ & $\bigcirc$ & $\bigcirc$ \\
WhatsApp & $\bigcirc$ & $\bigcirc$ & $\bigcirc$ & $\bigcirc$ & $\bigcirc$ & $\bigcirc$ & $\bigcirc$ \\
Instagram & $\bigcirc$ & $\bigcirc$ & $\bigcirc$ & $\bigcirc$ & $\bigcirc$ & $\bigcirc$ & $\bigcirc$ \\
TikTok & $\bigcirc$ & $\bigcirc$ & $\bigcirc$ & $\bigcirc$ & $\bigcirc$ & $\bigcirc$ & $\bigcirc$ \\
WeChat & $\bigcirc$ & $\bigcirc$ & $\bigcirc$ & $\bigcirc$ & $\bigcirc$ & $\bigcirc$ & $\bigcirc$ \\
Telegram & $\bigcirc$ & $\bigcirc$ & $\bigcirc$ & $\bigcirc$ & $\bigcirc$ & $\bigcirc$ & $\bigcirc$ \\
Snapchat & $\bigcirc$ & $\bigcirc$ & $\bigcirc$ & $\bigcirc$ & $\bigcirc$ & $\bigcirc$ & $\bigcirc$ \\
Reddit & $\bigcirc$ & $\bigcirc$ & $\bigcirc$ & $\bigcirc$ & $\bigcirc$ & $\bigcirc$ & $\bigcirc$ \\
X & $\bigcirc$ & $\bigcirc$ & $\bigcirc$ & $\bigcirc$ & $\bigcirc$ & $\bigcirc$ & $\bigcirc$ \\
Pinterest & $\bigcirc$ & $\bigcirc$ & $\bigcirc$ & $\bigcirc$ & $\bigcirc$ & $\bigcirc$ & $\bigcirc$ \\
LinkedIn & $\bigcirc$ & $\bigcirc$ & $\bigcirc$ & $\bigcirc$ & $\bigcirc$ & $\bigcirc$ & $\bigcirc$ \\
Other (Please Specify) & $\bigcirc$ & $\bigcirc$ & $\bigcirc$ & $\bigcirc$ & $\bigcirc$ & $\bigcirc$ & $\bullet$ \\
\hline
\end{tabular}
}
\end{table}

\item Have you ever posted transliterated content?
\begin{itemize}
    \item[$\bigcirc$] Yes \hspace{1cm} $\bigcirc$ No
\end{itemize}

\item Have you ever interacted with transliterated content?
\begin{itemize}
    \item[$\bigcirc$] Yes \hspace{1cm} $\bigcirc$ No
\end{itemize}
\end{itemize}

\subsection{Comprehension, Privacy Risks, and Offensive Content Detection}
\textit{Bangla speakers got one Bangla, one English, and one Banglish post and all the questions of this module thrice (once/text). For Hindi speakers, the same thing happened with Hindi, English, and Hinglish. Bangla-Hindi bilingual speakers got these questions five times for Bangla, Hindi, English, Banglish and Hinglish.}\\

Read the following (language based on selection) text -
\begin{itemize}

\item How frequently do you post transliterated content?
\begin{itemize}
    \item[$\bigcirc$] 1 - Never
    \item[$\bigcirc$] 2 - Rarely (a few times a month)
    \item[$\bigcirc$] 3 - Sometimes (a few times a week)
    \item[$\bigcirc$] 4 - Often (almost daily)
    \item[$\bigcirc$] 5 - Very Frequently (multiple times a day)
\end{itemize}

\item How frequently do you interact with transliterated content?
\begin{itemize}
    \item[$\bigcirc$] 1 - Never
    \item[$\bigcirc$] 2 - Rarely (a few times a month)
    \item[$\bigcirc$] 3 - Sometimes (a few times a week)
    \item[$\bigcirc$] 4 - Often (almost daily)
    \item[$\bigcirc$] 5 - Very Frequently (multiple times a day)
\end{itemize}

\item Do you understand the meaning of this post?
\begin{itemize}
    \item[$\bigcirc$] 1 - Not at all
    \item[$\bigcirc$] 2 - Slightly
    \item[$\bigcirc$] 3 - Somewhat
    \item[$\bigcirc$] 4 - Mostly
    \item[$\bigcirc$] 5 - Completely
\end{itemize}

\item Do you think this post is harmful?
\begin{itemize}
    \item[$\bigcirc$] Yes \hspace{1cm} $\bigcirc$ No \hspace{1cm} $\bigcirc$ Unsure
\end{itemize}

\item What action you would take on this post?
\begin{table}[htbp]
\centering
\resizebox{\linewidth}{!}{%
\begin{tabular}{l c c c c c}
\hline
\textbf{Action} & 1 - Very Unlikely & 2 - Unlikely & 3 - Neutral & 4 - Likely & 5 - Very Likely \\
\hline
Ignore it & $\bigcirc$ & $\bigcirc$ & $\bigcirc$ & $\bigcirc$ & $\bigcirc$ \\
React to it & $\bigcirc$ & $\bigcirc$ & $\bigcirc$ & $\bigcirc$ & $\bigcirc$ \\
Comment on it & $\bigcirc$ & $\bigcirc$ & $\bigcirc$ & $\bigcirc$ & $\bigcirc$ \\
Share it & $\bigcirc$ & $\bigcirc$ & $\bigcirc$ & $\bigcirc$ & $\bigcirc$ \\
Report it & $\bigcirc$ & $\bigcirc$ & $\bigcirc$ & $\bigcirc$ & $\bigcirc$ \\
\hline
\end{tabular}
}
\end{table}

\item Why would you report? (Open-ended)

\item Do you think this post contains privacy violation or security risks?
\begin{itemize}
    \item[$\bigcirc$] Yes \hspace{1cm} $\bigcirc$ No \hspace{1cm} $\bigcirc$ Unsure
\end{itemize}

\item Would you feel safe interacting with it?
\begin{itemize}
    \item[$\bigcirc$] Yes \hspace{1cm} $\bigcirc$ No \hspace{1cm} $\bigcirc$ Unsure
\end{itemize}

\item Do you think transliteration makes it harder to recognize harmful content?
\begin{itemize}
    \item[$\bigcirc$] Yes \hspace{1cm} $\bigcirc$ No \hspace{1cm} $\bigcirc$ Unsure
\end{itemize}

\item How difficult is it for automated moderation systems to detect such texts?
\begin{itemize}
    \item[$\bigcirc$] Easy \hspace{1cm} $\bigcirc$ Moderate \hspace{1cm} $\bigcirc$ Hard
\end{itemize}
\end{itemize}

\subsection{User-Perceived Impact on Digital Safety and Platform Moderation}
\begin{itemize}
\item Does transliteration make it harder to recognize harmful content?
\begin{itemize}
    \item[$\bigcirc$] Yes \hspace{1cm} $\bigcirc$ No \hspace{1cm} $\bigcirc$ Unsure
\end{itemize}


\item How concerned are you about multilingual hate speech?
\begin{itemize}
    \item[$\bigcirc$] 1 - Not at all concerned
    \item[$\bigcirc$] 2 - Slightly concerned
    \item[$\bigcirc$] 3 - Somewhat concerned
    \item[$\bigcirc$] 4 - Very concerned
    \item[$\bigcirc$] 5 - Extremely concerned
\end{itemize}

\item How concerned are you about multilingual misinformation?
\begin{itemize}
    \item[$\bigcirc$] 1 - Not at all concerned
    \item[$\bigcirc$] 2 - Slightly concerned
    \item[$\bigcirc$] 3 - Somewhat concerned
    \item[$\bigcirc$] 4 - Very concerned
    \item[$\bigcirc$] 5 - Extremely concerned
\end{itemize}

\item Have you experienced multilingual online harassment?
\begin{itemize}
    \item[$\bigcirc$] Yes \hspace{1cm}
    $\bigcirc$ No \hspace{1cm}
    $\bigcirc$ Prefer not to say
\end{itemize}



\end{itemize}





\end{document}